\documentclass{amsart}
\usepackage{amsmath}
\usepackage{amsfonts}
\usepackage{amssymb}
  \usepackage{paralist}
  \usepackage{graphics} %% add this and next lines if pictures should be in esp format
  \usepackage{epsfig} %For pictures: screened artwork should be set up with an 85 or 100 line screen
\usepackage{graphicx}  \usepackage{epstopdf}%This is to transfer .eps figure to .pdf figure; please compile your paper using PDFLeTex or PDFTeXify.
 \usepackage[colorlinks=true]{hyperref}
   \usepackage{tikz} 
\hypersetup{urlcolor=blue, citecolor=red}

\theoremstyle{definition}

\usepackage{xcolor}

\title[STV in CO House of Representatives]{Simulating Single Transferable Voting for the Colorado House of Representatives}

\author{Nora Nelson Laird}%\email{{n\_nelsonla2023}@coloradocollege.edu}
\author{Beth Malmskog}
\author{Iris Pixler}
\author{Katherine Rodbell}

\begin{document}
\maketitle

\begin{abstract}Social choice theory research demonstrates that single transferable voting (STV) results in more proportionally representative legislative bodies. We aim to understand how using multi-member districts and ranked ballots with STV would affect the representation of political parties in the Colorado House of Representatives. We investigated this objective by producing 10,000 multi-member districting plans of Colorado, generating ranked ballots for each of these plans using returns from the 2022 Colorado attorney general race, and simulating STV using these ballots. Our simulated STV elections for the Colorado House of Representatives gave more proportional representation for Democrats and Republicans than the current first-past-the-post system. Future research should explore how the implementation of STV would influence the representation of racial and language minority groups in the Colorado General Assembly to provide guidance on electoral reform in the state.\end{abstract}

\section{Introduction}\label{sec:intro} 
In a representative democracy, what voters have the most voice depends deeply on the system that is used to elect representatives.  First-past-the-post (FPTP) or plurality electoral systems with single-member geographic districts, in which the candidate with the most votes wins in each district, are simple and familiar to voters in the United States, primarily due to their widespread implementation. However, they have been critiqued for overrepresenting majorities, disadvantaging third parties, limiting voters' freedom of choice, and electing candidates who do not have a majority of support from their constituencies \cite{tideman1995single, caron1999end}. Further, outcomes in these systems are highly dependent on the boundaries that are drawn, creating an enormous potential for manipulation of and conflict over district borders. In response, researchers in the field of social choice theory have explored alternative voting systems that may mitigate these shortcomings, including single transferable voting (STV). STV is an electoral method that combines ranked choice voting and multi-member districts to elect candidates to a legislative body. Ranked choice voting (RCV) is a category of voting systems where voters rank candidates in order of preference. Multi-member districts are geographic areas where voters elect two or more representatives. STV has been implemented in public elections since 1856 and is currently used in Northern Ireland, Ireland, Malta, Australia, Scotland, New Zealand, Wales, and some US cities \cite{Hall2025, mohajan2011}. 

Though fairness in democracy and representation have long been studied by mathematicians, economists, and political scientists, the last 10-15 years have seen a flowering of mathematical, computational, and statistical methods in the area.  Computational power and insights have allowed researchers to analyze existing ranked choice voting data to see how frequently pathologies like monotonicity violations occur, and allowed advances in modeling ranked choice voting data \cite{benade2021ranked, mccune2024monotonicity, clellandAlaska}.  Ensemble analysis has been applied to study fairness and representation in districting plans (see \cite{herschlag2020quantifying, chen2015cutting, duchin2023redistricting, kekulandara2025partial} for just a few examples of a rich and rapidly developing area).  Researchers have studied both partisan and demographic dimensions of representation.  Several reserchers and groups, including the Data and Democracy Lab (DDL, formerly the MGGG Redistricting Lab), have made significant contributions to data science methodology and software for studying democracy and guiding electoral reform.  Some exciting recent work centers around RCV and multi-member districts, particularly racial representation, partially in response to threats to the federal Voting Rights Act \cite{benade2021ranked}. This work strengthens claims that STV is a more representative voting system due to its resistance to gerrymandering and its ability to account for both concentrated and dispersed communities of interest. We use several packages developed by DDL and the earlier MGGG, including Gerrychain, VoteKit, and maup \cite{mgggGerryChain, mgggVoteKit, mgggMaup}.  

The Colorado General Assembly is bicameral, consisting of the 65-member Colorado House of Representatives and the 35-member Colorado Senate. Colorado state representatives serve 2-year terms, and Colorado state senators serve staggered 4-year terms. Colorado has historically been considered a “purple” state, but has become more Democratic in recent years. So far, alternative voting systems have not been adopted in any state-wide or state legislature contests in Colorado. Most representatives in the state are elected through plurality elections. The few exceptions include the cities of Boulder, Fort Collins, Aspen, Telluride, and Basalt, which use RCV in some municipal elections, and Carbondale and Broomfield, which have voted for but not implemented ranked choice voting as of February 2026.  Proposition 131, a statewide ballot measure to adopt ranked choice voting, was rejected in November 2024,  with a vote of 53.5\% to 46.5\%. If passed, this measure would have replaced single-party primaries with all-candidate primaries and introduced RCV for general elections in Colorado \cite{Kenney}. Supporters of Proposition 131 claimed it would increase election competitiveness and make voters’ ballots more meaningful, while opponents were wary of the measure’s funding and how it could influence the voice of special interest groups \cite{CommonCauseCO2024, LeagueOfWomenVotersCO2024}.

Though 131 failed, Colorado voters and state legislators have enacted several innovative election measures. In 2018, Amendments Y and Z to the state constitution were passed by wide margins with bipartissan support, banning partisan gerrymandering and creating independent redistricting commissions to draw both congressional and state legislative boundaries.  The Colorado Voting Rights Act, signed May 12th, 2025, also has a wide potential impact on Colorado elections. It establishes several protections for Colorado voters, including heightened electioneering restrictions, prohibition of voter qualification based on gender identity or sexual orientation, and increased access to election information. Section 1-47-106 prohibits vote dilution—denying members of protected classes an equal opportunity to elect the candidates they support—and asserts that if election reforms exist that are likely to reduce vote dilution, they must be considered for implementation in Colorado \cite{COVRA}. It is important to note that vote dilution claims are generally proven by showing that it is possible to draw a majority-minority district (a district where a majority of the voting-age population belongs to a minority group), that members of this minority group vote together, and that they vote differently from members of the majority group \cite{gingles}. This is difficult to do in regions where the majority and minority groups are not geographically segregated, even though these regions may still have underrepresentative legislative bodies, something proportionally representative voting systems—like STV—may be able to address.

In this paper, we use the Colorado State House of Representatives as a case study in the potential of STV to provide more proportional representation than the current single-member district plurality system.  We focus on the partisan dimension of representation.  We generate 10,000 valid multi-member districting plans using MCMC methods, then generate simulated RCV votes for each plan using vote records from 2022 state-wide elections using a Plackett-Luce model and ballot truncation based on existing ranked choice data.  We then run elections on the simulated data in each district using STV.  We find that STV produces nearly proportional outcomes (with respect to party) and also performs well with respect to the Combined Support Metric, a more sophisticated measure of voter support for election outcomes.  Since RCV ballots and STV elections are computationally expensive to implement, many studies in the literature use ballot pools that are much smaller than the number of votes cast in historical plurality elections.  In contrast, our work uses full-scale simulations.  Our work is also distinctive in its focus on Colorado data and our custom model of ballot truncation.

{\bf Acknowledgements}  All authors of this work were partially supported by National Science Foundation Award DMS-2137661.  Authors Nelson Laird, Pixler, and Rodbell were also supported by Colorado College's Summer Collaborative Research Experience.  Author Malmskog was supported by Watkins funding from the Colorado College Department of Mathematics and Computer Science. 

\section{Background}\label{sec:background}

\subsection{Single Transferrable Vote}\label{subsec:STV}
STV was first developed in the 1800s as a system to give more proportional representation while avoiding some of the pitfalls of party-list proportional representation systems, and continues to be refined and studied today (see \cite{tideman1995single} for an excellent history and overview). In STV elections, each voter casts one ranked ballot, and their vote may be transferred among their most preferred candidates, as described below.
\begin{enumerate}
\item A vote threshold that a candidate needs to exceed to be elected is calculated. In this paper, we will use the Droop quota, $q$, where $q = \lceil \frac{\text{total votes}}{\text{available seats} +1} \rceil$.
\item If any candidate meets the quota, they are elected, and their surplus votes are redistributed proportionally based on voters’ next most preferred candidates.
\item If no candidate meets the quota, the candidate with the fewest first-place votes is eliminated, and their votes are redistributed as described in step 2.
\item This process is repeated until the required number of candidates has been elected.
\end{enumerate}
Although STV generally results in more proportional representation, it does, in rare cases, violate the monotonicity criterion of fair elections \cite{mccune2024monotonicity}. In other words, under STV, decreased support for a candidate can improve their chances of getting elected, while increased support can worsen their chances of getting elected. A monotonicity violation can be a cause for concern when it provides avenues for strategic voting, allowing voters to reach more favorable outcomes by casting ballots that do not represent their actual preference order of candidates. However, the complexity of the STV algorithm makes it unlikely for any one voter to take advantage of this anomaly \cite{mccune2024monotonicity}. Furthermore, Arrow’s Impossibility Theorem states that every RCV system violates some reasonable fairness criterion, indicating that rare monotonicity violations do not provide sufficient reasoning to discount STV as an acceptable and even preferable method \cite{black1969arrow}.

\subsection{Modeling STV}

One challenge in studying STV is the relative lack of RCV data from real elections. Data bright spots include Scotland, which has RCV data from local elections going back to 2007, and Cambridge, MA, which has elected city council members by STV since 1941. Simulating STV in any place that currently votes by plurality requires that researchers first simulate RCV data, often by employing a theoretical or empirical model for voter behavior and making many modeling assumptions.  There are several established models for voter behavior that have been developed and implemented by various researchers over the years.  The software package VoteKit brings together implementations of many existing and some new models for ranked choice ballot generation, as well as an integrated implementation of the STV election protocol \cite{benade2021ranked}. We employ this implementation in our work.

In each model, {\bf blocs} are defined as disjoint sets of voters with similar voting patterns. Some examples include blocs determined by race or political party. Each bloc has a corresponding {\bf slate}, a disjoint set of candidates preferred by members of that bloc. {\bf Cohesion parameters} are defined for each bloc, reflecting how likely voters in that bloc are to vote for each slate.  Candidate strength parameters, known as {\bf preference intervals}, are set for each bloc and consist of values for each candidate based on their relative popularity among members of that bloc. If there are $n$  blocs with $n$ corresponding slates, each bloc has $n$ preference intervals and $n$ cohesion parameters, one for each slate. 

Consider two blocs, $A$ and $B$, with corresponding $A$ and $B$ slates, each with defined candidates. A user of the models would set four cohesion values, $\pi_{AA}$, $\pi_{AB}$, $\pi_{BB}$, and $\pi_{BA}$, representing how likely members of bloc $A$ are to vote for slate $A$, members of bloc $A$ are to vote for slate $B$, members of bloc $B$ are to vote for slate $B$, and members of bloc $B$ are to vote for slate $A$, respectively. For each bloc, the sum of the cohesion parameters must equal one. For example, for bloc $A$, $\pi_{AA}+\pi_{AB}=1$. 

Similarly, there are four preference intervals, representing bloc $A$’s support for each slate $A$ candidate, bloc $A$’s support for each slate $B$ candidate, bloc $B$’s support for each slate $B$ candidate, and bloc $B$’s support for each slate $A$ candidate, which we will name $V_{AA}$,  $V_{AB}$, $V_{BB}$, $V_{BA}$ respectively. These are represented below as vectors:
 $V_{AA} = \langle v^A_{A_1}, v^A_{A_2}, \cdots, v^A_{A_k}\rangle$
$V_{AB} = \langle v^A_{B_1}, v^A_{B_2}, \cdots, v^A_{B_l}\rangle$
 $V_{BB} = \langle v^B_{B_1}, v^B_{B_2}, \cdots,v^B_{B_l} \rangle$
 $V_{BA} = \langle v^B_{A_1}, v^B_{A_2}, \cdots, v^B_{A_k} \rangle$
where $k$ and $l$ are the number of candidates in slate $A$ and $B$, respectively. Each element in a vector represents the support a candidate is receiving from that bloc. For example, the element $v^A_{A_1}$ in $V_{AA}$ is the proportion of support from bloc $A$ voters towards the first candidate in the $A$ slate. The values in each preference interval must sum to one. So, for instance, summing the values in $V_{AA}$ gives $v^{A}_{A_1}+ v^A_{A_2}+ \cdots+v^A_{A_k}=1$.

The numbers in each preference interval can be fixed in the model or determined probabilistically. In the latter case, a set of values, one for each candidate in the preference interval, is drawn from a symmetric Dirichlet distribution, which assigns probabilities to groups of values that sum to one. For the case of $V_{AA}$, these probabilities are given by 
\[P( v^A_{A_1}, v^A_{A_2}, \cdots, v^A_{A_k}) = B \prod_i (v^A_{A_i})^{\alpha-1},\] 
where $B$ is a normalizing constant and $\alpha$ is a measure of concentration chosen by the user of the ballot generators. As $\alpha$ approaches infinity, the chances of having a preferred candidate among voters in a bloc increase; as $\alpha$ approaches 0, support is more likely evenly divided between the candidates; and when $\alpha = 1$, every set of weights is equally likely. In the literature, values of 0.5 and 2 have been used to simulate cases with even support and a strong candidate, respectively.  An $\alpha$ value must be set for each possible pair of one slate and one bloc. So, in the case of blocs $A$ and $B$, we would have four $\alpha$ values: $\alpha_{AA}, \alpha_{AB}, \alpha_{BB}, \alpha_{BA}$, each dictating how values will be drawn for $V_{AA}$, $V_{AB}$, $V_{BB}$, and $V_{BA}$, respectively.

Blocs, slates, cohesion parameters, and preference intervals determine how ballots are generated in the following four models, all of which are implemented in VoteKit.
\begin{enumerate}
\item The \textbf{Plackett-Luce} (PL) method creates a ballot one ranking at a time, filling spots on the ballot based on bloc cohesion and preference intervals \cite{plackett1975analysis, luce1959individual}. This method is described as modeling an “impulsive voter,” because it simulates a voter filling out each slot on the ballot independently of their other decisions on the ballot. There are two versions of this model: 
\begin{enumerate}
    \item {\bf Name-Plackett-Luce (n-PL)} is a version in which ballots are filled out by candidate name. To create a ballot for a voter, say a voter in bloc $A$, each bloc $A$ preference interval is rescaled by its corresponding cohesion parameter. Then, all the rescaled bloc $A$ preference intervals are concatenated to create a new preference interval. This interval forms a distribution, which is drawn from without replacement to fill out the ballot.  
    \item {\bf Slate-Plackett-Luce (s-PL)} is a version in which ballots are first filled out by slate, and then candidates are filled in based on the slate ranking. In this version, the cohesion parameters determine the probability that a slate will be selected on a ballot. For example, $\pi_{AA}$ is the probability that slate $A$ will be selected for a slot on the ballot of a bloc $A$ voter. Once a slate is on the ballot as many times as there are candidates in that slate, it is removed from the possible choices. Next, with a complete ballot filled out by slates, candidates are selected from those slates, drawing from the preference intervals. So, for instance, for an $A$ slate spot on a bloc $A$ voter’s ballot, a candidate will be drawn from a distribution of the values in $V_{AA}$.
\end{enumerate}
\item The {\bf Bradley-Terry (BT)} method creates ballots based on pairwise comparisons \cite{bradley1952rank}. This method is described as modeling a “deliberate voter,” because it simulates a voter basing each ranking on their other choices. There are two versions of this model:
\begin{enumerate}
    \item {\bf Name-Bradley-Terry (n-BT)} is a version in which ballots are filled out by pairwise comparisons of candidates. It starts off the same as n-PL, creating a new preference interval. The probability of a ballot being chosen is calculated using the values in this updated preference interval. Let $C_i$ represent candidate $i$’s value in this preference interval. As an example, the probability that the ballot selected for a voter ranks candidate 1 first, candidate 2 second, and candidate 3 third ($C_1>C_2>C_3$) is proportional to $P(C_1>C_2)P(C_2>C_3)P(C_1>C_3)$, where $P(C_1>C_2) = \frac{C_1}{( C_1+C_2)}$. 
    Each voter’s ballot is then assigned according to these probabilities.
    
    \item {\bf Slate-Bradley-Terry (s-BT)} is a version in which ballots are filled out by pairwise comparisons of slates. To create a ballot for a voter, say a voter in bloc $A$, we start by filling out slate labels only. The probability of a slate ordering is calculated using cohesion parameters. As an example, the probability that the slate ordering selected for a voter ranks a candidate from slate $A$ first, a candidate from slate $B$ second, and a candidate from slate $A$ third  $(A > B > A)$ is proportional to $\pi_{AA}(\pi_{AB})$. Each voter’s slate ordering is assigned according to these probabilities, and their candidates are filled in using their preference intervals.
\end{enumerate}
\item {\bf The Alternating Crossover (AC)} method  \cite{mgggLowell, mgggYakima} creates ballots by dividing voters into two groups: bloc voters and crossover voters. Bloc voters rank all the candidates in their slate above all the candidates in the opposing slate. Crossover voters rank a candidate from the opposing slate first, and then alternate between candidates from their slate and the opposing slate until candidates from one of the slates run out. Note that bloc voter ballots will always be complete (all candidates are ranked), but crossover ballots may be incomplete (not all candidates are ranked) if the number of candidates in each slate is not the same. The number of voters in each group is determined by the cohesion parameters. For example, $\pi_{AA}$ is the proportion of bloc voters from bloc $A$ and $\pi_{AB}$ is the proportion of crossover voters in bloc $A$. Once the slates are set on the ballot based on voters’ groups, candidates from the corresponding slates are drawn from the preference interval distributions. 

\item {\bf The Cambridge Sampler (CS)} method \cite{benade2021ranked} creates ballots based on historical voting data from city council elections in Cambridge, MA, one of the few US cities that has consistently used STV in public elections since its implementation in 1941 \cite{ProportionalRepresentationFoundation}. Under this method, voters are split into four groups: bloc A voters ranking a slate A candidate first, bloc A voters ranking a slate B candidate first, bloc B voters ranking a slate B candidate first, and bloc B voters ranking a slate A candidate first. Based on their first choice, each voter is randomly assigned a ballot from a distribution of Cambridge ballot types. The blocs used by the Cambridge Sampler are defined by race, where bloc A consists of POC voters, slate A consists of candidates preferred by POC voters, bloc B consists of white voters, and slate B consists of candidates preferred by white voters. Because Cambridge ballots are not all complete, the ballots generated using the CS can also be incomplete.

\end{enumerate}

In testing the four different models while varying parameters (preference intervals, number of candidates, number of seats, location, election, etc.), the DDL found that, on average, all four models produced similar, highly proportional outcomes \cite{benade2021ranked}. They noted that the CS method seemed to deviate the most from the others, which they hypothesized was due to more variation in ballot lengths. 

\subsection{Combined Support Metric}\label{sec:CSM}
Furthermore, to evaluate the effectiveness of STV, Benade et al. \cite{benade2021ranked} wanted to compare their simulation results to a benchmark based on voter support rather than one based on the number of voters in each bloc. Since members from each bloc who voted for candidates of the opposing bloc would be satisfied if those candidates were elected, they calculated the expected seat share based on the proportion of voters in each bloc and that bloc’s cohesion parameter. For example, to determine seat share for bloc $A$ based on the combined support metric, they take the proportion of bloc $A$ voters ($P_A$), multiply it by the proportion of those voters who vote in line with their party, ($\pi_{AA}$), then add the proportion of bloc $B$ voters ($P_B$) multiplied by the proportion of those voters who do not align with their party ($\pi_{BA}$), to get a proportion of voters who support slate $A$. Letting $S_A$ represent the bloc $A$ seat share expected based on combined support, we have the equation:
\[S_A \approx P_A \cdot \pi_{AA} + P_B \cdot \pi_{BA}.\]

Using this metric, the authors \cite{benade2021ranked} reported that running STV simulations on ballots generated by all four models across different scenarios produced election results that were in line with their predicted seat share.

\subsection{Ballot Truncation}
One limitation of PL and BT is that they do not allow for incomplete ballots. Hoffman et al. expand on the work done by the DDL by investigating ballot length \cite{hoffman2021proportionality}. In elections with ranked ballots, not every ballot is filled out completely—in fact, often only a small fraction are. Some voters may only list candidates from their party, some may only list their top few, and some may list just a single candidate (also known as a bullet vote). For example, in Cambridge city council elections from 2007 to 2017, fewer than 4\% of voters submitted a complete ranking \cite{hoffman2021proportionality}. The authors explore three possible ways ballots generated from PL and BT can be truncated. One of these methods, independent truncation, uses ranked ballot data from Cambridge to create a distribution of likely ballot lengths. Each generated ballot is assigned a length based on this distribution, with the hope of creating ballots that better represent real-world scenarios. 

\subsection{Ensembles}
Ensemble analysis is a statistical technique that has, in the last 15 years, become increasingly important in the world of redistricting (see \cite{duchin2022political} for a survey and context). In this technique, an ensemble of thousands (or even millions) of legally valid redistricting plans is randomly generated.  Then, voting data from real elections is overlaid on the maps and the partisan balance of each district is computed, creating a statistical profile of the properties of maps drawn without partisan bias.  If a proposed or enacted district map is very unusual compared to this ensemble, this may be evidence of partisan bias in that map.  In our context, we will use an ensemble of randomly-generated multi-member districting plans to ensure that our results are robust with respect to the choice of districting plan.

One way these districting plans are created is through the ReCom method \cite{deford2021recombination}. In this method, a dual graph is formed in which geographic areas (precincts, census blocks, or, in our case, existing State House districts) are associated with vertices, and edges are placed between vertices corresponding to areas that share a border. A districting plan then corresponds to a graph partition.  Given an initial districting plan, ReCom creates new plans by the following Markov Chain process:
\begin{enumerate}
\item A random pair of neighboring districts is selected and the districts are merged. 
\item A random spanning tree of the dual graph of the merged district is generated, and a single edge is removed from the spanning tree to create two subgraphs (where the edge is chosed to balance population), each one representing a new district. 
\end{enumerate}
Over the course of many steps, a large ensemble of possible districting plans is created. 

Researchers have previously used ensemble analysis to evaluate Colorado's 2011 and 2021 congressional and state legislative redistricting plans \cite{clelland2022colorado, Clelland2021a, Clelland2021b, Clelland2021c}. They generated millions of possible plans to serve as a baseline for assessing seat share, competitiveness, and minority representation, and found no compelling evidence that the enacted plans were drawn with bias or effective partisan manipulation.

In 2021, University of Colorado Boulder undergraduate student Catherine Brennan explored how multi-member districts with STV could minimize gerrymandering in Colorado \cite{brennan}. She found, using a simplified STV model in an ensemble of multi-member districts for the state house, that multi-member districts resulted in less variation in seat share than single-member districts. This result indicates that multi-member districts are less prone to gerrymandering, due to the decreased likelihood of outliers and extreme results. Furthermore, she found using her model that STV produced more proportional seat share outcomes based on party. 

\section{Methods}

We seek to understand how using multi-member districts and ranked ballots with STV would affect the representation of the two major political parties in the Colorado House of Representatives. We investigate this objective by generating ranked ballots using returns from the 2022 Colorado Attorney General race, truncating those ballots, and simulating STV elections for 10,000 multi-member district maps of Colorado.  In this section, we present the details of our methods.

\subsection{Initial Model}
\begin{figure}[h]
\centering
\includegraphics[width=0.9\textwidth]{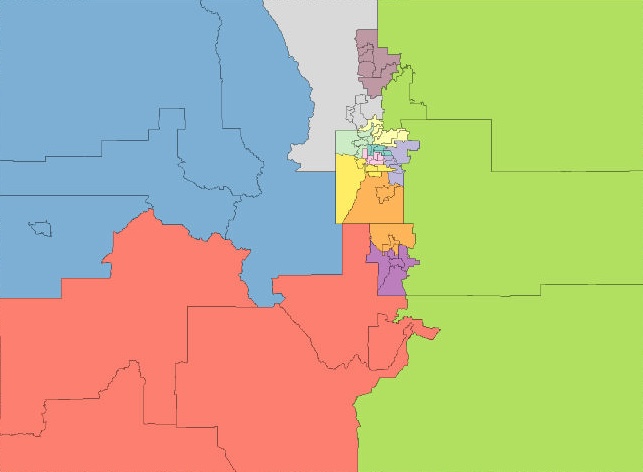}
\caption{Our initial multi-member district map, created using data from the U.S. Census Bureau and ArcGIS (U.S. Census Bureau, 2024).}\label{fig1}
\end{figure}

We first combined Colorado’s 65 state house districts into 13 multi-member districts (Figure \ref{fig1}), each of which would elect five representatives. Each multi-member district consists of five existing districts, grouped based on geographical adjacency and federal congressional districts. By being conscious of existing federal congressional districts, which were drawn using input from communities of interest, we were better able to keep culturally and geographically distinct areas together.

Our first approach to generating ballots was writing a Python script that takes returns from Colorado elections and outputs ballots where all candidates of a voter’s party are ranked above candidates of the other party. After this proof of concept, we went on to consider the four more complex models of ballot generation available in VoteKit \cite{mgggVoteKit}. We decided not to use the CS method because it is based on racial demographics as opposed to party. Additionally, we felt that the ballots created by PL and BT were more realistic than those created by AC. We ultimately chose s-PL because we thought impulsive voter behavior that prioritizes party over candidate best represented real-world voters.

In order to simulate elections in Colorado, we needed benchmarks for the number of Democratic and Republican voters. We considered using party registration data, but ultimately decided to use previous election results to incorporate the opinions of unaffiliated voters, of which there are many in Colorado, partially due to its semi-open presidential primaries. 

The first elections we used were the 2024 Colorado House of Representatives races (all election results mentioned in this paper were taken from the Colorado Secretary of State’s Historical Election Database, which can be accessed at \url{https://historicalelectiondata.coloradosos.gov/}.)T We determined the number of Democratic and Republican voters for each multi-member district by summing the vote totals by party for each of its single-member districts. For our first test, we set both parties to be completely cohesive, candidate strength equal for all candidates, and five candidates (one per single district) for each of the two major parties. We then used VoteKit’s s-PL ballot generation function to create a ballot for each voter. Simulating the elections using VoteKit’s STV function with these ballots resulted in 35 Democrats and 30 Republicans being elected, which is proportional to the number of voters from each party in our data set.

\subsection{Exploration for Final Model}
To construct our final model, we wanted to see how varying different parameters affected the outcomes of simulated STV elections. Letting $D$ and $R$ represent the Democrat and Republican blocs, respectively, we have cohesion parameters $\pi_{DD}$,  $\pi_{DR}$, $\pi_{RR}$, $\pi_{RD}$ and preference intervals $V_{DD}$, $V_{DR}$, $V_{RR}$, $V_{RD}$ made using alpha values $\alpha_{DD}$,  $\alpha_{DR}$,  $\alpha_{RR}$, $\alpha_{RD}$. The tests we ran are outlined below.
\begin{enumerate}
    \item {\bf Candidate Strength Parameter:} We ran several tests with varying preference intervals and high bloc cohesion ($\pi_{DD}=1$, $\pi_{DR}=0$, $\pi_{RR}=1$,  $\pi_{RD}=0$). We did this by changing the alpha values, using one-half, one, and two, the numbers used by the DDL to make preference intervals where each candidate has roughly the same strength, preference intervals where there is an equal chance of any breakdown of candidate strength, and preference intervals where one candidate is stronger than the other candidates. We focused mainly on varying each bloc’s preference intervals for their own slate ($\alpha_{DD}$ and $\alpha_{RR}$), keeping the each bloc’s preference interval for the opposing slate completely random, setting $\alpha_{DR}$, $\alpha_{RD}$ equal to one.  Our tests include both parties having a strong candidate within their party ($\alpha_{DD} = 2$, $\alpha_{RR}=2$), neither party having a strong candidate within their party ($\alpha_{DD} = 0.5$, $\alpha_{RR}=0.5$), and each party having a strong candidate while the other did not ($\alpha_{DD} = 0.5$, $\alpha_{RR}=2$ and $\alpha_{DD} = 2$, $\alpha_{RR}=0.5$). Our results did not vary from test to test and were all proportional, indicating that STV is resilient to relative candidate strength when bloc cohesion is very high. 

    \item {\bf Cohesion Parameter:} Varying cohesion, we ran four tests: high cohesion within both parties ($\pi_{DD} = 0.9$, $\pi_{DR} = 0.1$, $\pi_{RR} = 0.9$, $\pi_{RD} = 0.1$), low cohesion within both parties ($\pi_{DD} = 0.6$, $\pi_{DR} = 0.4$, $\pi_{RR} = 0.6$, $\pi_{RD} = 0.4$), and low cohesion in one party with high cohesion in the other ($\pi_{DD} = 0.9$, $\pi_{DR} = 0.1$, $\pi_{RR} = 0.6$, $\pi_{RD} = 0.4$ and $\pi_{DD} = 0.6$, $\pi_{DR} = 0.4$, $\pi_{RR} = 0.9$, $\pi_{RD} = 0.1$). We maintained an equal strength for all candidates in each test. From our tests, we found that when both groups had similar levels of cohesion, the results were proportional, but when one group had higher cohesion, they unsurprisingly performed much better. 

    \item {\bf Candidate Strength and Cohesion Parameters Together:}  By varying candidate strength and bloc cohesion together, we found that having a strong candidate can result in a party receiving fewer seats under STV if the other party has evenly divided support. In an extreme case we tested—where Democrats had low cohesion and a strong candidate while Republicans had high cohesion and equally preferred candidates—Republicans were able to win a narrow majority of state house seats, something we did not see in any of our other tests. Although candidate strength and bloc cohesion have a stronger influence on outcomes when varied together than they do individually, our results did not vary much across these tests unless we used exaggerated cohesion values.

    \item {\bf Experimenting with Election Choices:} Using our original model but changing the election from which we drew our voter set (2022 Attorney General, 2022 Governor, 2022 State Treasurer, 2022 Secretary of State, and 2024 US Senator) resulted in proportional outcomes even as the number of voters per party changed. To obtain these voter sets, we combined vote totals from precincts to calculate the total number of Democratic and Republican voters in each multi-member district.

    \item {\bf Experimenting with Third Parties:} To better understand how STV influences the success of third-party candidates, we experimented with adding a third-party bloc to our model. We used the 2022 Colorado gubernatorial race, this time including the voters for the Libertarian candidate. Because each multi-member district has a small share of Libertarian voters (around 1\%), Libertarian candidates were rarely elected in our simulation, a reasonable proportional outcome. However, when we adjusted the cohesion parameters in an extreme example so that Libertarians always voted together (100\% cohesion) and major party voters had some crossover (80\% cohesion each), a few Libertarian candidates were elected. 

\end{enumerate}

\subsection{Final Model}
After observing the effects of changing different parameters, we chose the following to implement in our final model. The code for our final model and the files it uses can be found here: \url{https://github.com/katherinerodbell/Summer-2025-Colorado-STV-Research}.
\begin{enumerate}
    \item {\bf Number of Candidates:} First, we considered the number of candidates to run per party. We noted that many Colorado House of Representatives elections in 2022 had uncontested candidates. Thus, we thought it would be unlikely for every district to run an equal number of Republican and Democratic candidates, especially in districts where one party had a much higher vote share than the other. Therefore, we wanted to base the number of candidates on the proportion of voters from each party. Since having a large number of candidates from one party can divide voters and hurt the party’s chances of securing seats, we speculated that the optimal number of candidates a party would run is the number of seats they expect to win proportionally, plus one, giving them the opportunity to secure one more seat than expected. This method left us with seven candidates. In order to allow for more extreme possibilities, we decided to have nine candidates per district, with the number of candidates per party determined by the number of seats they would expect to win proportionally, plus two. This also takes into account the fact that political parties do not have complete control over the candidates who end up on the ballot, and may end up running more than the optimal number.

    \item {\bf Candidate Strength:} Next, we considered candidate strength. Due to the numerous factors that determine candidates’ popularity, we expected that preference intervals would vary across multi-member districts. There could be cases where certain candidates are very well-liked or disliked. On the other hand, we could also imagine cases in which all candidates had comparable popularity. To account for these different scenarios, we decided to make preference intervals random, setting all alpha values in each multi-district equal to one. 

    \item {\bf Cohesion:} To settle on how likely a member of each political party is to vote for candidates in their party, we used exit poll data from a sample of 2022 and 2024 Colorado elections, including federal Senate, Presidential, and Gubernatorial \cite{ColoradoPollingInstitute2024, COExitPolls2022, COExitPolls2024}. Averaging the percentage of self-reported Democrats and Republicans who voted in line with their party, we found that Democrats voted for Democratic candidates approximately 93\% of the time, and Republicans voted for Republican candidates approximately 88\% of the time. So we set $\pi_{DD} = 0.93$, $\pi_{DR} = 0.07$, $\pi_{RR} = 0.88$, and $\pi_{RD} = 0.12$. Looking at unaffiliated voters, we found that the percentage voting Democrat and Republican remained similar across races. Thus, we assumed that unaffiliated voters tend to vote cohesively with a certain party, roughly in line with the high cohesion values we set.  

    \item {\bf Election Data:} Lastly, we chose to use the 2022 Attorney General election for our voter set. To get a reasonable count, we considered three factors when selecting the election: year, prominence of candidates, and voter turnout. We only considered statewide elections that occurred after state redistricting in 2021, which included the 2022 Governor, Secretary of State, State Treasurer, and Attorney General races. Next, we wanted to minimize the possibility of a well-known candidate changing race outcomes. Therefore, we ruled out the most recent Governor and Secretary of State elections, since there were strong public opinions on both winners. This left us between the State Treasurer and Attorney General elections. We assumed candidates in those races were less known, and thus, voters would be more likely to vote along party lines. Since voter turnout was higher in the Attorney General election, we decided to use that race for our model. 

\end{enumerate}

We then used the parameters above to generate ballots using the s-PL model. 

In the implemented s-PL model, all candidates are ranked on all ballots. We wanted to account for ballots of different lengths in our model. Based on the methods of \cite{hoffman2021proportionality}, we created a distribution of ballot lengths to apply to our ballots. We created our distribution by adapting ballot length data from the Cambridge 2019 city council election. Although using data from Cambridge is standard in RCV literature, it is not a perfect representation of how STV would look in Colorado. Cambridge city council elections are non-partisan and have used STV long enough to educate the electorate. Voter education is important in the implementation of alternative voting systems, as demonstrated by the 2024 city council election in Portland, Oregon, where an informational campaign instructing voters to rank all candidates was reflected in ballot lengths \cite{mgggPortland}.

After considering multiple years of Cambridge data, we decided to base our model on the 2019 election because it was the most recent for which ballots were available on the Ranked Choice Voting Resource Center website (\url{https://www.rcvresources.org/}), and had a maximum ballot length of fifteen, which was most similar to our maximum ballot length of nine. Since the number of possible ballot lengths is different, we could not directly use the Cambridge distribution and had to adapt it to our simulation. 

\begin{figure}
   \centering
    \includegraphics[width=\linewidth]{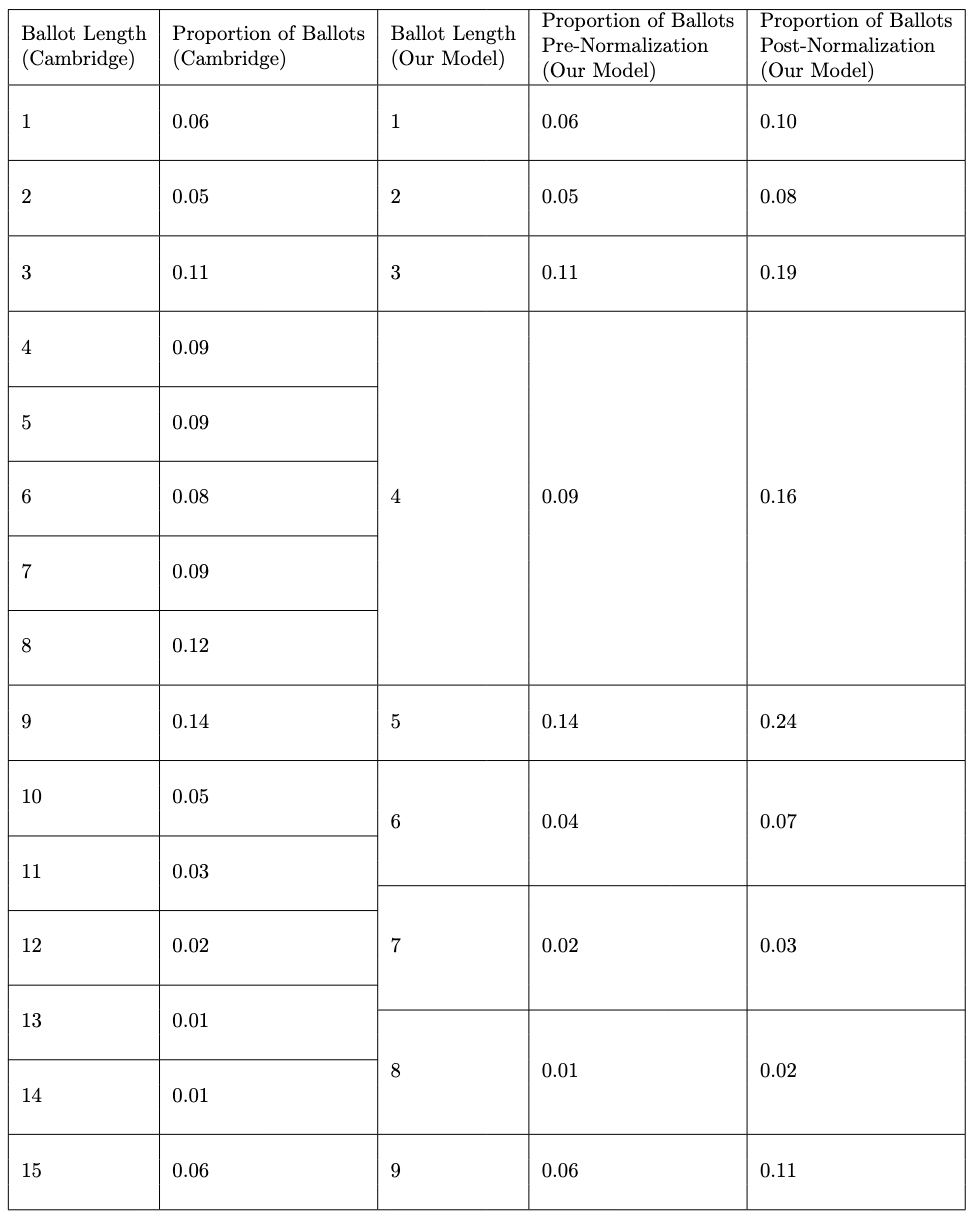}
    \caption{This table shows the process we went through to adapt the Cambridge 2019 city council election ballot length distribution to our simulation parameters. The first and second columns show the proportion of each ballot length in the Cambridge distribution. The third and fourth columns show how we adjusted the Cambridge distribution’s fifteen possible ballot lengths to represent our nine possible ballot lengths. The fifth column shows the final percentages we used in our ballot length distribution. All numbers in this table are rounded to two decimal places.}
    \label{fig:fig2}
\end{figure}

Across multiple Cambridge elections, we saw peaks at ballot lengths of one, three, the number of seats available (in the case of Cambridge, nine), and the maximum ballot length. We transferred the proportion of ballots at each of these peaks to their corresponding points in our distribution. Finally, we evenly distributed the proportions between the key points of the Cambridge distribution to their respective ballot lengths in between the key points in our distribution. This process is described in more detail below and represented in Figure \ref{fig:fig2}.  
\begin{enumerate}
    \item We kept the proportions of ballots of lengths one, two, and three the same in our distribution. 
    \item We averaged out the proportions of ballots of lengths four through eight in the Cambridge distribution to get the proportion of ballots of length four in our distribution. This resulted in a proportion of  $\frac{(0.09+0.09+0.08+0.09+0.12)}{5} \approx 0.09$ for ballots of length four in our distribution. 
    \item We used the proportion of ballots of length nine in the Cambridge distribution as the proportion of ballots of length five in our distribution because these numbers represent the number of available seats in Cambridge and our model, respectively. 
    \item We split the proportion of ballots of lengths ten through fourteen in the Cambridge distribution across the proportion of ballots of lengths six through eight in our distribution. We transferred the Cambridge proportions of lengths ten, twelve, and fourteen directly to our lengths six, seven, and eight, respectively. To also include lengths eleven and thirteen from the Cambridge distribution, we divided the proportion for those two ballot lengths into thirds, and distributed the six thirds among lengths six, seven, and eight in our distribution, with six getting two-thirds of eleven, seven getting one-third of eleven and one-third of thirteen, and eight getting two-thirds of thirteen. Finally, to standardize, we multiplied our three resulting proportions by three-fifths. So, for the proportion of ballot length 6 in our distribution, we get $\frac{3}{5}(0.05+\frac{2}{3}(0.03)) \approx 0.04$; for the proportion of length seven, we get $\frac{3}{5}(0.02+\frac{1}{3}(0.03)+\frac{1}{3}(0.01)) \approx 0.02$; and for the proportion of length eight, we get $\frac{3}{5}(0.01+\frac{2}{3}(0.01)) \approx 0.01$.
    \item We set the proportion of our maximum ballot length (nine) equal to the proportion of the maximum ballot length (fifteen) in Cambridge.
    
\end{enumerate}

\begin{figure}[h]
\centering
\includegraphics[width=0.5\textwidth]{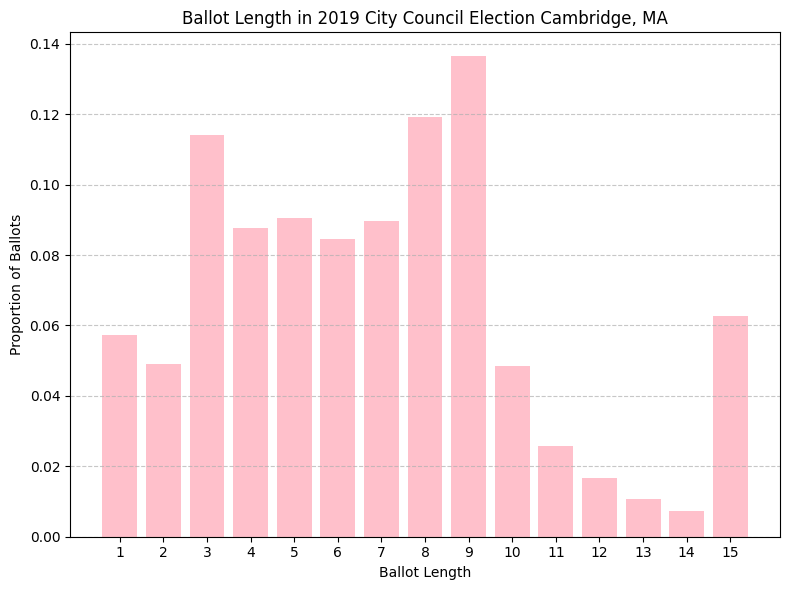}\includegraphics[width=0.5\textwidth]{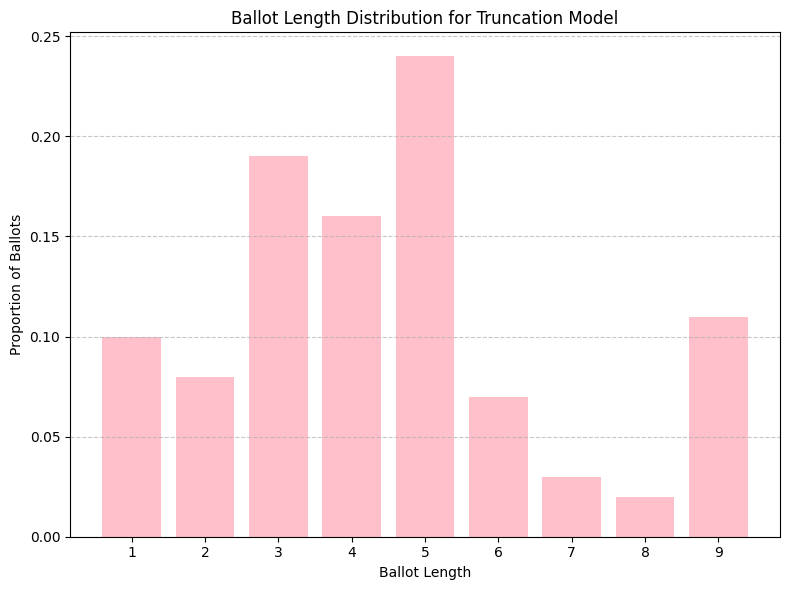}
\caption{Using data from the Ranked Choice Voting Resource Center, we created a distribution of ballot lengths from the 2019 City Council election in Cambridge, MA (left). We condensed this distribution from a maximum ballot length of fifteen to a maximum ballot length of nine to sample from when truncating the ballots in our model (right).
}\label{fig:fig3}
\end{figure}

This process gave us the general shape of the Cambridge distribution. We then normalized our proportions (using the normalizing constant 1.7266) to obtain our final distribution, shown to the right of the Cambridge distribution in Figure \ref{fig:fig3}.

%truncation distributions here

We went through each ballot generated in our model and assigned it a length drawn from our distribution. We then truncated each ballot to its assigned length before using the ballots to simulate STV elections.

\subsection{Ensemble}\label{sec:ensemble}
We decided to employ ensemble analysis in our study to reduce the influence of the qualities of any individual map (such as our initial plan, used in our first model). Section 47 and Amendments Y and Z of the Colorado constitution require that legislative districts be contiguous, reasonably compact, have approximately equal populations, preserve communities of interest, keep municipalities together when possible, maximize politically competitive districts, and that they do not favor political parties or dilute the influence of minority groups \cite{COConstAmendY, COConstAmendZ}. When creating our sample of maps, we sought to uphold these requirements as much as possible. 

We chose to build the ensemble of multi-member district maps by combining current state house districts, instead of precincts, as described in Section \ref{sec:ensemble}. Since the current districts were drawn to meet existing requirements, we assumed that combining them into multi-member districts would create maps that also likely meet these requirements. However, to be cautious, we also incorporated GerryChain’s built-in contiguity, compactness, and population deviation constraints to ensure our districting plans were reasonable. We set the population deviation to 2\%, stricter than the current 5\% constraint in Colorado, because our multi-member districts have larger populations than the current districts \cite{COConstArtVSec46}. We used the maup Python package \cite{mgggMaup} to fix overlaps and gaps in the Colorado Independent Redistricting Commission’s 2021 final approved house plan shapefile \cite{COApprovedHousePlan}, allowing us to establish our dual graph. 

\begin{figure}[h]
\centering
\includegraphics[width=1.00\textwidth]{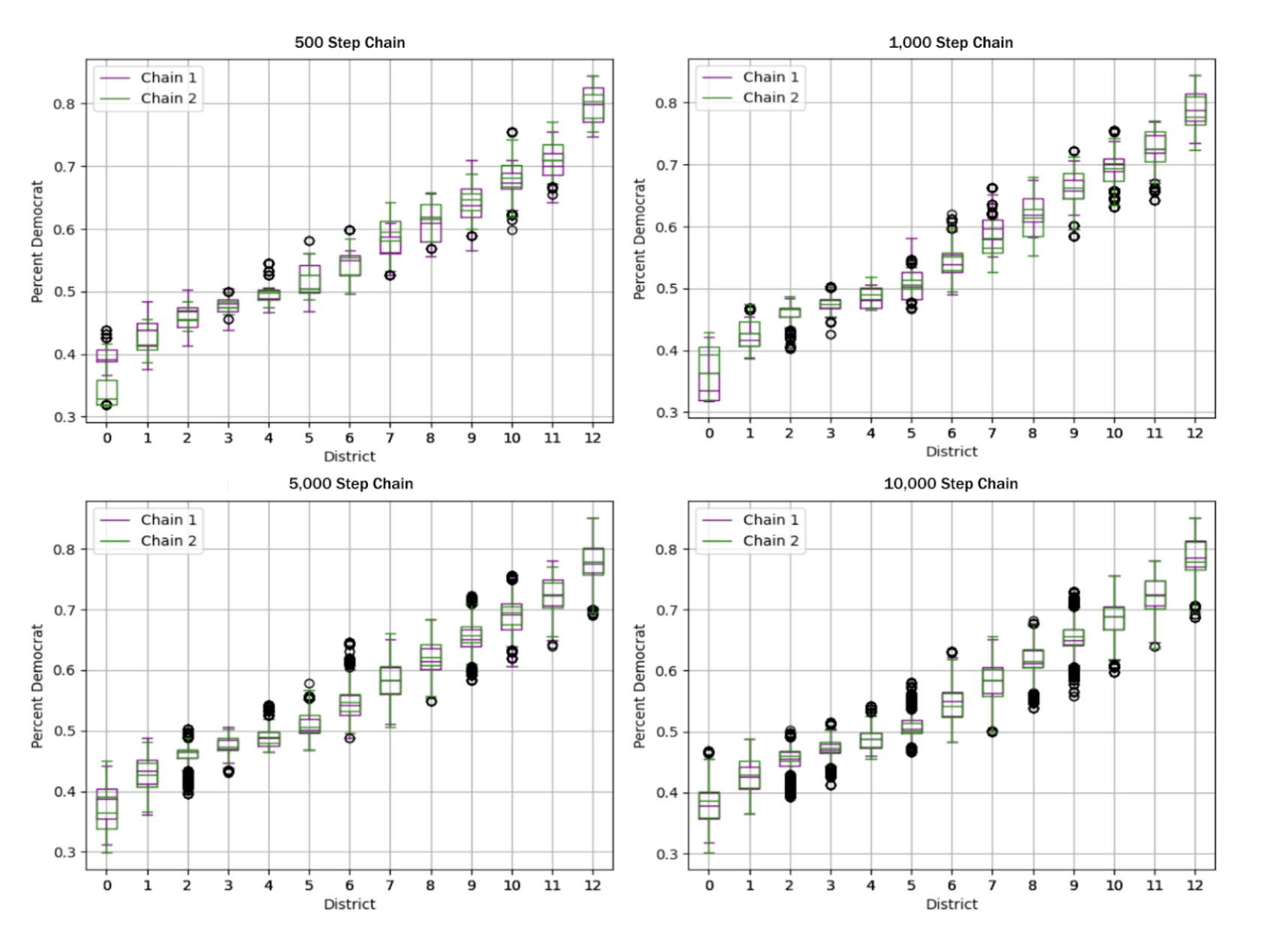}
\caption{These four boxplots display the proportion of Democratic voters from the least Democratic district (0) to the most Democratic district (12) for two different Markov chains of the same length (500, 1,000, 5,000, and 10,000). To demonstrate that a chain of a given length produces sufficient mixing, we want to show that the starting point of said chain does not influence its results. In other words, the boxplots of two different chains of the same length should be visually indistinguishable. Above, the purple and green boxplots vary for chains of lengths 500, 1,000, and 10,000, but are nearly identical for the chains of length 10,000.
}\label{fig:fig4}
\end{figure}

\begin{figure}[h]
\centering
\includegraphics[width=1.00\textwidth]{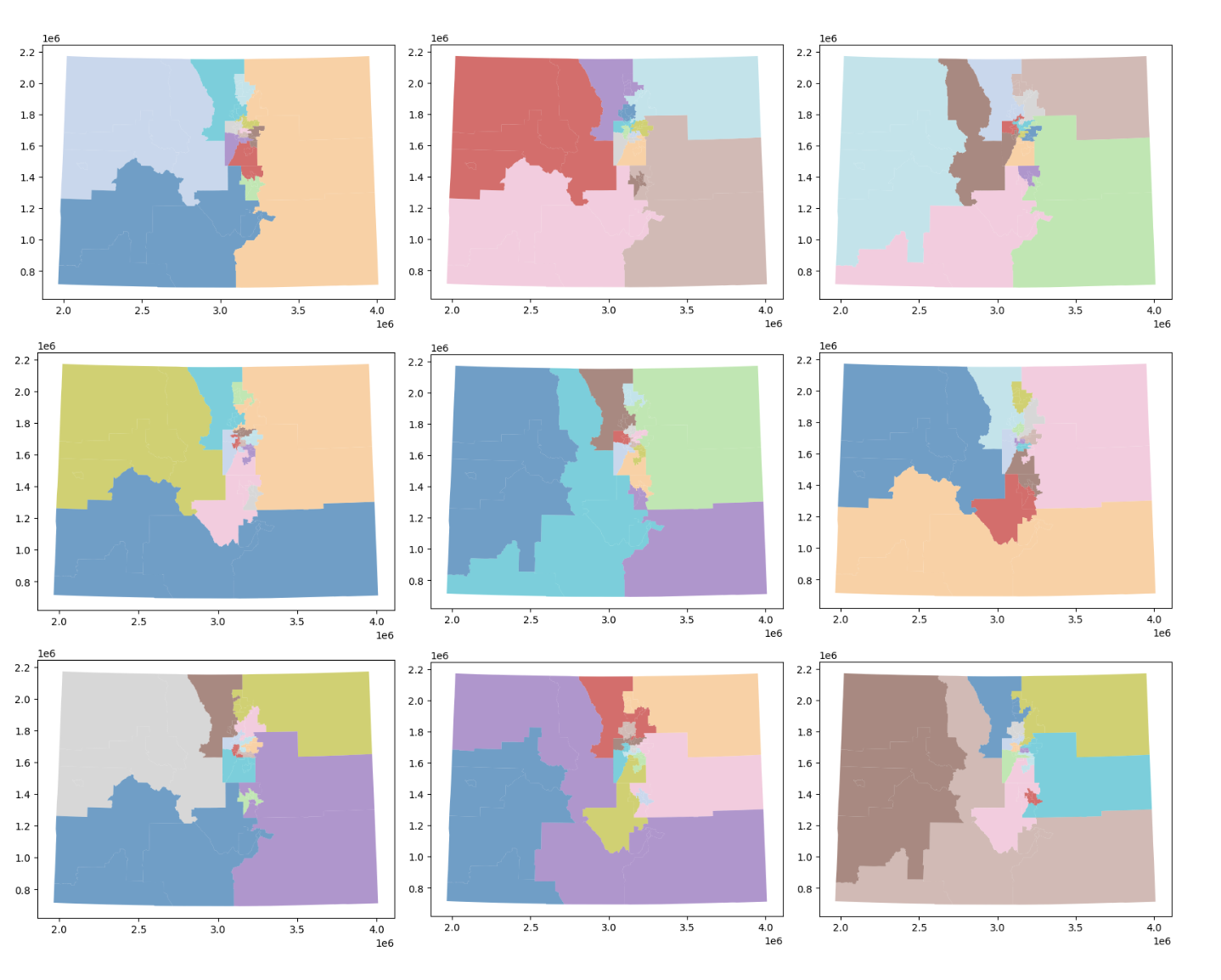}
\caption{Nine Colorado multi-member district plans generated by ReCom and selected from different points in the Markov chain.
}\label{fig:fig5}
\end{figure}

When deciding the length of a Markov chain for ensemble analysis, it is important to ensure that it demonstrates effective mixing. In other words, the starting point of the chain should not influence the results. We tested chains of lengths 500, 1,000, 5,000, and 10,000 by running two of each and overlaying their boxplots displaying Democratic seat share. The boxplots of two chains that are long enough to be completely mixed will be visually indistinguishable from one another. When we overlaid the boxplots for chains of lengths 500, 1,000, and 5,000, we found that there was enough variation to indicate that they were not sufficiently mixed. When we overlaid the boxplots for the chain of length 10,000, there was little variation, suggesting that a chain of this length presents sufficient mixing for the variable we are trying to measure, seat share by party. The overlaid boxplots from chains with 500, 1,000, 5,000, and 10,000 steps are shown in Figure \ref{fig:fig4}. An example of several maps from different points in a 10,000-step chain are shown in Figure \ref{fig:fig5}.

%box plots figure here

%3x3 of sample maps here

\section{Results and Discussion}

\subsection{Simulation Results and Analysis}
\begin{figure}[h]
\centering
\includegraphics[width=0.5\textwidth]{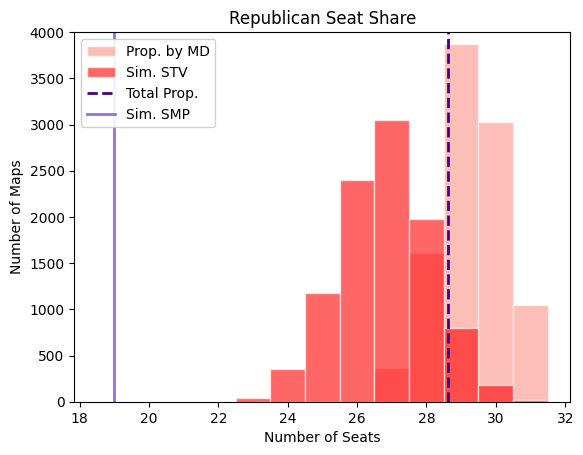}\includegraphics[width=0.5\textwidth]{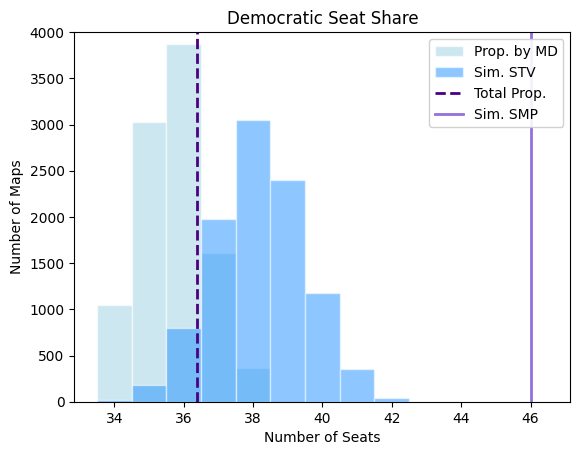}
\caption{The darker red and darker blue histograms represent the distribution of seat shares for Republicans and Democrats, respectively, from the STV simulations on our 10,000 maps. The light red and light blue histograms represent the distribution of expected proportional seat share by multi-member district for each map. The dashed dark purple lines represent proportional seat share for each party based on the proportion of Democratic and Republican voters in the state (from the 2022 Attorney General Election). The light purple lines represent the number of districts in which each party would have the majority in a plurality election in the enacted single-member districts with the Attorney General election data, which was the same as the composition of the Colorado House of Representatives in 2022, 46 Democrats and 19 Republicans.
}\label{fig:fig6}
\end{figure}

We ran STV elections using our generated ballots for the 10,000 multi-member district maps of Colorado from our Markov chain. For every map, along with the simulated election results, we recorded the number of seats both parties would win if each multi-member district election had proportional outcomes. In Figure \ref{fig:fig6}, the distribution of simulated STV seat-share outcomes is shown in darker red and darker blue, while the distribution of proportional outcomes is shown in light red and light blue for Republicans and Democrats, respectively. It is important to note that the same conclusions could be drawn from a histogram displaying seat share for either party, but we have included both so as not to show partiality.  

The respective means from the distributions of proportional outcomes were approximately 35.7 seats for Democrats and 29.3 for Republicans. From our STV simulations, the average Democratic seat share was approximately 38.2, and the average Republican seat share was approximately 26.8. 

As seen in Figure \ref{fig:fig6}, the proportional distributions and simulated STV distributions overlap, with a difference of 2.5 seats between the means. The light purple lines representing the number of seats that each party would have won by plurality in single-member districts under the given election data (which was the same as the Colorado House of Representatives party composition in 2022) lie outside of the simulated STV distributions, with a 7.8-seat difference between these lines and the means of the distributions. These results suggest that using multi-member districts and ranked ballots with STV produces more proportional results than using single-member districts with plurality for the Colorado House of Representatives. 

We believe that Democrats maintain a slight advantage in our simulation compared to the proportional results due to the difference in cohesion values. The smaller cohesion value measured for our Republican bloc gives Democrats more seats than proportionally expected. 

%'poster results' here (figure 6)

\begin{figure}[h]
\centering
\includegraphics[width=0.5\textwidth]{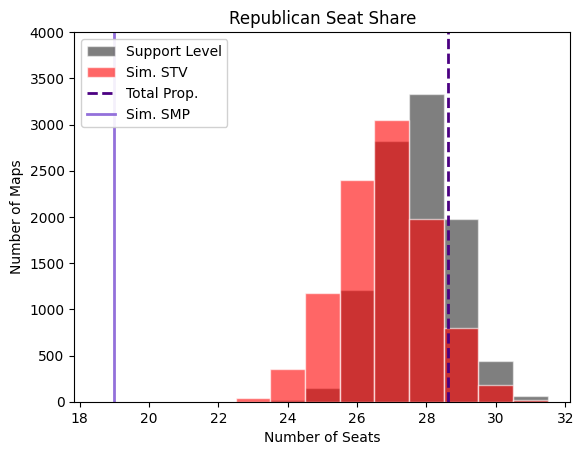}\includegraphics[width=0.5\textwidth]{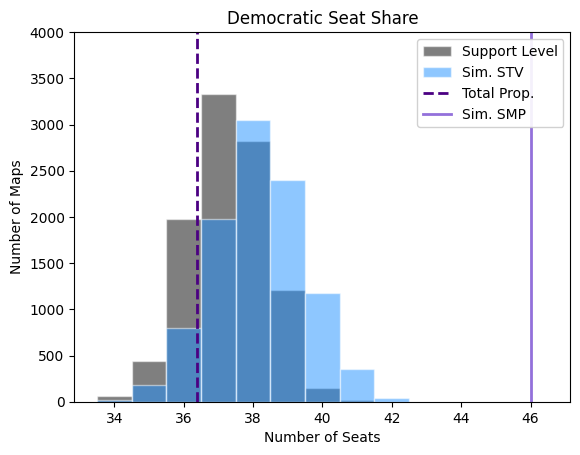}
\caption{The red and blue histograms represent the distribution of seat shares for Republicans and Democrats, respectively, from the STV simulations on our 10,000 maps. The gray histograms represent the distribution of seat share using the combined support metric from \cite{benade2021ranked}. The light purple lines represent the number of districts in which each party would have the majority in a plurality election in the enacted single-member districts with the Attorney General election data, which was the same as the composition of the Colorado House of Representatives in 2022, 46 Democrats and 19 Republicans.
}\label{fig:fig7}
\end{figure}

For each of the 10,000 maps, we applied the combined support metric from \cite{benade2021ranked} as another comparison point for our results (see Section \ref{sec:CSM}). We calculated this metric for each multi-member district and combined the results to get the statewide seat-share for each map. These combined support distributions are displayed in gray in Figure \ref{fig:fig7}, while the STV simulation distribution is still displayed in red and blue. 

The difference in the mean seat share for the combined support metric and our simulation is 0.9, which is notably less than the difference of means from Figure \ref{fig:fig6}. This affirms that the election results are in line with voters’ preferences. 

%'gray backing' plots here 

%'red/blue' table outside of figure
\begin{table}[h]
    \centering
    \begin{tabular}{|l|l|l|l|l|}
         \hline & \begin{tabular}[c]{@{}l@{}}
         Avg. Prop. by \\ Multi-Member District \end{tabular} & Avg. Sim. STV & \begin{tabular}[c]{@{}l@{}}Avg. Combined \\ Support Metric\end{tabular} & Sim. SMP \\ \hline
         Dem. Seat Share & 35.7 & 38.2 & 37.3& 46\\ \hline
        Rep. Seat Share & 29.3 & 26.8 & 27.7 & 19 \\ \hline
    \end{tabular}
    \caption{This table can be used to quickly compare the average seat share for Democrats and Republicans based on proportional results by multi-member district, our simulation, the combined support metric from \cite{benade2021ranked}, and the actual seat share in 2022.}\label{tab1}%
\end{table}

\begin{figure}[h]
\centering
\includegraphics[width=0.5\textwidth]{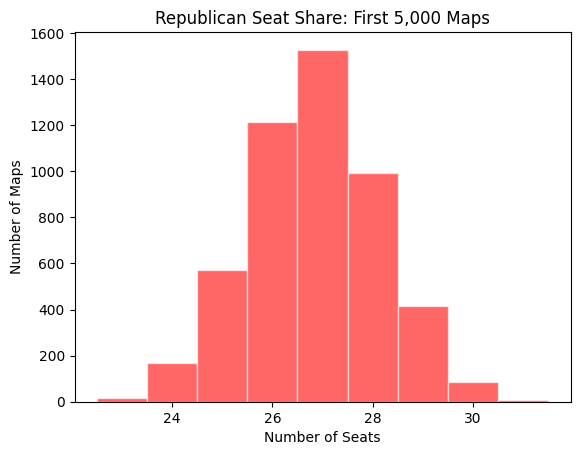}\includegraphics[width=0.5\textwidth]{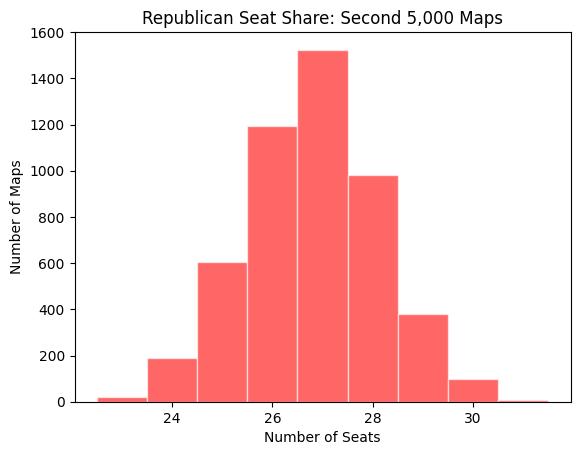}
\includegraphics[width=0.5\textwidth]{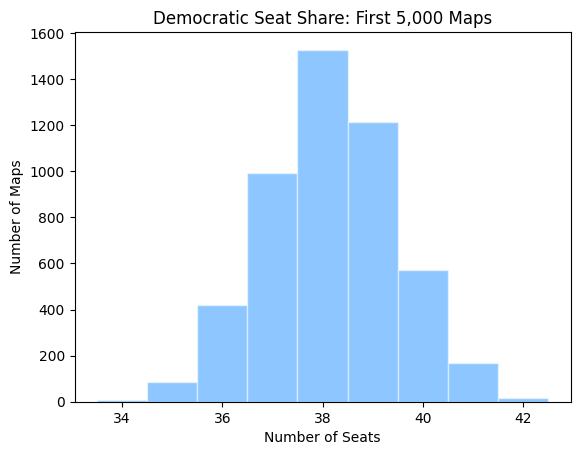}\includegraphics[width=0.5\textwidth]{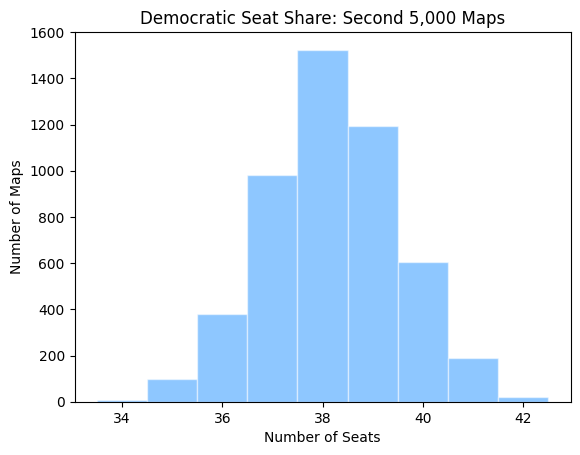}
\caption{These histograms compare the seat share for Democrats (shown in blue) and Republicans (shown in red) from our first and second 5,000 maps.
}\label{fig:fig9}
\end{figure}

Additionally, to further verify that a Markov chain of length 10,000 was sufficient for mixing, we compared the seat share results from the first 5,000 maps generated by our chain with those from the second 5,000 maps. We found that the seat share distributions are nearly identical for the first 5,000 maps, starting at map 1, and the second 5,000 maps, starting at map 5,001 (see Figure \ref{fig:fig9}), and the box plots for two different chains of 10,000 steps are nearly indistinguishable.

%first 5000 vs second 5000 here

\subsection{Comparison to Previous Work}
The claim that STV provides more proportional representation, affirmed many times in the literature, is upheld in our findings. However, we were surprised by how alike our results were to those of an ensemble of single-member plurality districting plans in \cite{Clelland2021b}. The spread of seat share distributions they found—using single-member districts and plurality (SMP)—is equal to the spread of our seat share distributions, with both ranging from 23 to 30 seats for Republicans and 35 to 42 seats for Democrats. Democrats performed slightly better in their analysis, with a mean of 38.39 seats compared to our mean of 38.2 seats. However, it's important to note that very different election data was used in these studies--the 2021 single-member district study used a composite of eight elections from 2016-2020, while our histograms were based on the single 2022 Attorney General election.  Further, the 2021 study used precincts as building blocks, while our study used enacted State House districts.  Thus, we cannot draw any strong conclusions by comparing our work and \cite{Clelland2021b}.  It would be interesting to do a more directly comparable ensemble analysis on single- and multi-member districts. This would also allow for more understanding of how wide the range of vote shares might vary under STV versus SMP.  In \cite{brennan}, a simplified model of STV (approximate proportionality for each district) produced a much narrower range of likely vote shares across an ensemble of multi-member districting plans versus an ensemble of single-member districting plans for the Colorado State Senate.  Would this still hold up with our modeling choices, i.e., simulating full-scale elections in the Colorado State House?  The fact that the spread of vote shares were approximately the same in our analysis and that of \cite{Clelland2021b} (though not with the same election data) makes this an important question for further investigation.

\section{Conclusion}

Our work presents outcomes for simulating STV elections for the Colorado House of  Representatives. We have found evidence to suggest that STV elections with multi-member districts and RCV result in more proportional party representation than the current elections with single-member districts and plurality voting; however, our methodology has several limitations that warrant discussion:
\begin{enumerate}
    \item The lack of existing RCV data forced us to generate simplified ballots for use in our simulation. Bloc cohesion and candidate strength parameters are difficult to ascertain with confidence, and real ranked ballots are influenced by a number of complex factors, including, but not limited to, party strategy, trends in ballot length connected to voter demographics, and the number of candidates running. This lack of real-world data also manifested itself when we generated our multi-district maps, as we could not accurately predict how redistricting laws, such as population deviation, would be influenced by larger districts.
    \item Simulating STV elections for each map took an average of 20 minutes, resulting in a total computation time of 200,000 minutes (almost 140 days) for all 10,000 maps. We ran 5,000 elections simultaneously, utilizing ten computers each running ten Python scripts, bringing our run time to 2,000 minutes—a much shorter but still considerable amount of time. Although simulating STV on more maps would have produced more robust results, we were limited to 10,000 by time and the quantity and strength of computers available to us. 
    \item Because we decided to use only the Republican and Democratic voters from the 2022 Colorado Attorney General race to estimate the number of voters per multi-member district, our simulations do not include voters who cast ballots for third-party candidates. Since relatively few voters cast ballots for candidates outside the Democratic and Republican parties in the state-wide races we considered, this has likely not had a large effect on this simulation, but research suggests RCV increases support for third-party candidates, meaning that our results are likely not representative/predictive of what would occur in actual State House elections conducted using STV \cite{simmons2022ranked}.

\end{enumerate}

Future research could address the limitations of this study by varying the parameters that we used in our simulation, including the type of ballot generation model, election, cohesion parameters, preference intervals, and number of candidates. Specifically, we would like to better understand how chosen alpha values influence the proportionality of seat share distribution outcomes. One way this could be done is by keeping alpha values consistent across maps and multi-member districts, for example, setting all alpha values to 0.5 or all alpha values to 2. Furthermore, our ensemble analysis could potentially be improved upon by using precincts as building blocks (as opposed to the current state house districts) and preserving county and municipality boundaries in our districting plans. We would also like to conduct a full robustness study, varying modeling parameters like candidate strength, cohesion, choice of election, and ballot truncation both in isolation and combination in subsequent work.

An important extension of this work would incorporate a study of racially polarized voting in Colorado and use racial and language minority groups as blocs instead of political parties. We expect that, in line with our party analysis, STV would result in more proportional racial and language minority representation in the Colorado House of Representatives. If this analysis provides robust results showing that the current first-past-the-post system dilutes the votes of members of protected classes, it could be an avenue for electoral reform under the Colorado Voting Rights Act as discussed in Section \ref{sec:intro}.

\bibliographystyle{plain}

%\bibliography{STV_bib.bib}

\end{document}